# Unveiling the relative timing jitter in counter propagating all normal dispersion (CANDi) dual-comb fiber laser

**NEERAJ PRAKASH,**[1] **SHU-WEI HUANG**[,2] **AND BOWEN LI**[1,3]

*Department of Electrical, Computer, and Energy Engineering, University of Colorado, Boulder, Colorado 80309, USA*
[1]*These authors contribute equally*
[2] *ShuWei.Huang@colorado.edu*
[3] *Bowen.Li@colorado.edu*

**Abstract:** Counter-propagating all-normal dispersion (CANDi) fiber laser is an emerging high-energy single-cavity dual-comb laser source. Its relative timing jitter (RTJ), a critical parameter for dual-comb timing precision and spectral resolution, has not been comprehensively investigated. In this paper, we enhance the state-of-the-art CANDi fiber laser pulse energy from 1 nJ to 8 nJ. We then introduce a novel reference-free RTJ characterization technique that provides shot-to-shot measurement capability at femtosecond precision for the first time. The measurement noise floor reaches $1.6\times10^{-7}$ $fs^2/Hz$, and the corresponding integrated measurement precision is only 1.8 fs [1 kHz, 20 MHz]. With this new characterization tool, we are able to study the physical origin of CANDi laser's RTJ in detail. We first verify that the cavity length fluctuation does not contribute to the RTJ. Then we measure the integrated RTJ to be 39 fs [1 kHz, 20 MHz] and identify the pump relative intensity noise (RIN) to be the dominant factor responsible for it. In particular, pump RIN is coupled to the RTJ through the Gordon-Haus effect. Finally, solutions to reduce the free-running CANDi laser's RTJ are discussed. This work provides a general guideline to improve the performance of compact single-cavity dual-comb systems like CANDi laser benefitting various dual-comb applications.

## 1. Introduction

Dual-comb systems, which comprise of a pair of optical frequency combs (OFC) with slightly different repetition rates, are attracting much research attention due to their applications in diverse fields including ranging [1], rotation sensing [2], spectroscopy [3] and asynchronous sampling [4]. One of the most important features required to implement a high-performance dual-comb system is the relative stability, such as relative timing jitter between the two frequency combs, which directly impacts the precision and resolution in all dual-comb applications. Even though actively stabilized dual-comb systems can exhibit outstanding low-noise performances [5], [6], their inevitably high cost and complexity have pushed the interest towards other alternatives. An emerging trend to passively maintain the relative stability focuses on generating the OFC pair in a single laser cavity [4], [7]–[14]. In such implementations, since the two pulses share the same cavity, there is inherent common-mode noise rejection (CMNR), which eliminates the requirement for sophisticated phase lock loops (PLLs) [15].

So far, single-cavity dual-comb systems have been realized in various laser architectures including solid state lasers [7], [9], fiber lasers [4], [10], [12]–[14] and on-chip waveguide [11] lasers. The single-cavity dual-comb fiber lasers are of particular interest due to its compact and robust nature. Therein, a promising approach to multiplex two OFCs in one fiber laser is bidirectional operation which simultaneously provides overlapping optical spectra and minimized crosstalk between the two OFCs [12]–[15]. In particular, we have demonstrated the first counter-propagating all-normal dispersion (CANDi) mode-locked fiber laser that broke through the energy limit of existing dual-comb fiber lasers by two orders of magnitude using all-normal dispersion cavity and nonlinear polarization rotation (NPR) mode-locking [14].

Energetic pulses with more than 1-nJ energy and flat broadband spectra were generated from both directions simultaneously, with a repetition-rate difference tunable from 0.1 Hz to 100 Hz. Pulse energies in nJ regime are critical for various nonlinear dual-comb applications such as dual-comb Raman spectroscopy and THz spectroscopy [4], [16]–[18]. For instance, 4-nJ pulses was required to realize coherent Raman spectro-imaging with a titanium-sapphire dual-comb system and 2 nJ was utilized to realize self-triggered asynchronous optical sampling THz spectroscopy with a single fiber dual comb laser [4], [16]. Few nJ dual-comb lasers have also been used as pumping sources in mid IR dual-comb OPO systems for dual comb spectroscopy [19], [20].

In this work, we enhance the state-of-the-art CANDi fiber laser pulse energy from 1 nJ to 8 nJ and introduce a novel reference-free relative timing jitter characterization technique that enables studies to gain deeper insights into the relative timing jitter characteristics of CANDi fiber laser. The technique combines the principles of dispersive Fourier transform (DFT) and spectral interferometry (SI) to measure the shot-to-shot relative timing jitter at femtosecond precision, enabling the comprehensive characterization of relative timing jitter up to the Nyquist frequency, or half the repetition rate. Currently, our technique achieves an ultra-low noise floor of $1.6\times10^{-7}$ fs$^2$/Hz and the corresponding integrated measurement precision is only 1.8 fs [1 kHz, 20 MHz]. Compared to the state-of-the-art relative timing jitter characterization technique using optical heterodyne detection [21], our measurement noise floor is 5 times better and no narrow-linewidth reference laser is required. We first verify that the cavity length fluctuation does not contribute to the relative timing jitter as its effect is fully suppressed by the CANDi's single-cavity design. Then we measure the integrated relative timing jitter to be 39 fs [1 kHz, 20 MHz] and we identify the pump relative intensity noise (RIN) to be the dominant factor responsible for CANDi fiber laser's relative timing jitter. In particular, pump RIN is coupled to the relative timing jitter through the Gordon-Haus effect. Finally, solutions to further reduce the free-running CANDi fiber laser's relative timing jitter have been discussed.

## 2. CANDi Laser with Enhanced Pulse Energy

Firstly, a new CANDi laser with further enhanced pulse energies is constructed. The laser structure is similar to the first demonstration [14]. However, the HI1060 fibers have been replaced with 10-μm core large mode area fibers. This simple modification has helped effectively in lowering the nonlinearity and thus pushing the highest obtainable pulse energy to 8 nJ in both laser directions. The fundamental repetition rate ($f_{rep}$) is 40 MHz and the repetition rate difference ($\Delta f_{rep}$) is tunable from 0.1 to 100 Hz. The laser schematics is shown in Fig. 1(a)

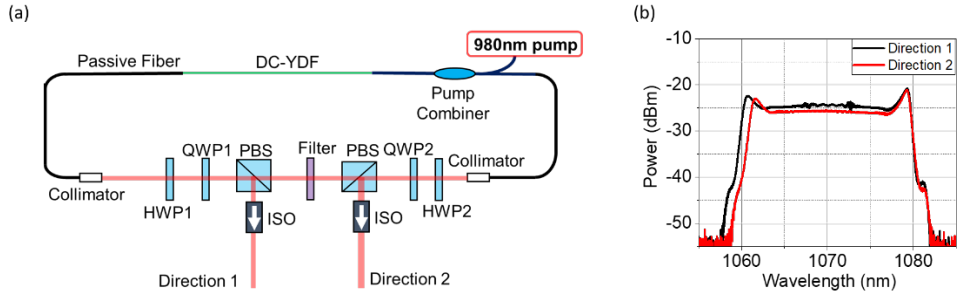

Fig. 1. (a) Experimental setup of the 8nJ CANDi laser. The gain fiber is 2 m of 6μm core double-cladding ytterbium-doped fiber (DC-YDF) and the passive fiber is 10 μm core large mode area (LMA) fiber. The pump combiner has double cladding LMA structure. The length of LMA fiber is 1.5 m on each side of DC-YDF. (b) Optical spectra of the mode-locked pulse in both directions. HWP: Half-wave plate, QWP: Quarter-wave plate, PBS: Polarizing beam splitter, ISO: Isolator

and the optical spectra is shown in Fig. 1(b).

## 3. Impact of Free-space Cavity Length Fluctuation

The impact of cavity length fluctuation on the relative timing jitter is then studied. In general, cavity length fluctuation is a major noise source of timing jitter in conventional OFCs and an effective way to suppress it is to actively control a movable mirror in the free-space section of the laser to compensate the optical path length variation of the laser cavity [22]. Here we study whether cavity length stabilization can suppress the relative timing jitter noise as well.

To study the impact of cavity length stabilization, we stabilize the $f_{rep}$ of one direction of the CANDi laser using a PLL and a slow piezo stage mounted under one of the fiber collimators. Since the locking bandwidth is ~25 Hz, we use frequency counters to measure the effect of locking in the low frequency regime. Fig. 2(a) shows the power spectral density (PSD) of the $f_{rep}$ of the individual laser directions in the free running configuration and when the $f_{rep}$ of direction 2 is locked to an RF source. When the laser is free running, the PSD of both the directions (black and red curve in Fig. 2 (a)) are overlapping. As shown in Fig. 2(a), locking one of the laser directions provides visible noise suppression in both directions. This is different from work by Link et al, where the authors observed a complete decoupling of $f_{rep}$ noise between the two OFCs caused by the semiconductor saturable absorber mirror [23]. In the CANDi laser, cavity length stabilization is a viable technique to improve timing jitter noise of both directions simultaneously. Meanwhile it is also obvious that the stabilization induces different level of noise suppression in two directions. The higher residual noise level in direction 1 compared to the locked direction are attributed to the uncommon mode noise, i.e., the relative timing jitter.

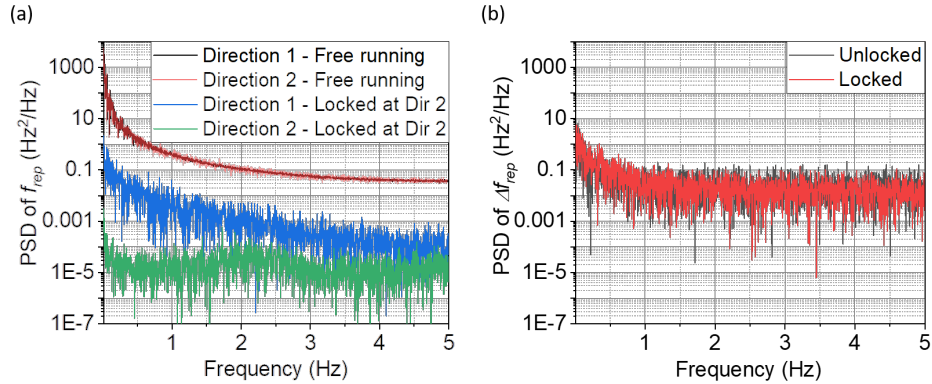

Fig. 2. (a) PSD of $f_{rep}$ of the individual directions of the CANDi laser for both free running case and when $f_{rep}$ of direction 2 is phase locked to a RF reference. (b) PSD of relative frequency measured by beating the 24th harmonics of both directions (scaled to fundamental repetition rate difference) for free running case and when $f_{rep}$ of direction 2 is phase locked to a RF reference. All the measurements are done using frequency counter.

To directly study the effect of cavity length fluctuation on the relative timing jitter noise, we measure it by mixing the 24th harmonics of the repetition rate of both the directions in a frequency mixer and measuring the beat note (i.e., 24th harmonics of repetition-rate difference $\Delta f_{rep}$) using a frequency counter. Fig. 2(b) shows the PSD of the beat note scaled to fundamental repetition rate difference with (red trace) and without (black trace) locking the $f_{rep}$ of direction 2. As observed, locking $f_{rep}$ of one direction have no effect on the relative timing jitter noise. This proves that the cavity length fluctuation impacts the two OFCs with exact same manner and therefore only affects the common mode noise and has negligible effect on the uncommon mode noise. Thus, free-space cavity length is a parameter that can be used to fine tune the average repetition rate without changing the repetition rate difference or relative timing jitter of CANDi.

**4. Relative Timing Jitter Measurement using Real-time Interferometry**

In order to identify the real noise source of relative timing jitter of CANDi laser or any other dual-comb systems and realize effective noise suppression accordingly, it is important to develop a technique for comprehensive relative timing jitter characterization up to the Nyquist frequency. However, it is extremely challenging since the single-cavity dual-comb fiber lasers typically exhibit only femtosecond (fs)-level relative timing jitter even under free-running conditions thanks to the inherent CMNR and the jitter noise is distributed in frequency domain up to tens of MHz. Hence, innovative measurement techniques that simultaneously achieve high timing resolution and high speed are the need of the hour to characterize the relative timing jitter of free running dual-combs. Conventional comb timing jitter measurement methods like balanced optical cross-correlation [24] and optical heterodyne [25] cannot be applied to dual-comb systems, where pulses temporally walk off from each other in only a few roundtrips. Recently, Shi et al., demonstrated a relative timing jitter characterization technique for dual-comb systems using asynchronous optical sampling (ASOPS). Sub-femtosecond measurement precision has been achieved, but only random-walk noise can be measured and the noise PSD is not available for analysis [26]. Another technique to measure relative timing jitter of dual-combs is using an indirect phase comparison between two fast photodetectors. However, the noise floor is relatively high ($10^{-5}$ fs$^2$/Hz) [27], which is too high for characterizing the fs level relative timing jitter in dual-comb fiber lasers. Recently, Sandro et al., demonstrated a high-resolution relative timing jitter characterization technique using optical heterodyne detection. Even though lower noise floor ($8\times10^{-7}$ fs$^2$/Hz) was realized, the implementation is more complex requiring additional narrow-linewidth single frequency lasers [21]. Here, we introduce a novel reference-free technique that do not require narrow-linewidth reference lasers and achieves 5 times lower noise floor to effectively study the relative timing jitter of dual-comb lasers using the DFT based real-time spectral interferometry [28]. While the heterodyne detection approaches provide the capability to measure relatively lower frequency noise, our method is a reference-free technique that is insensitive to intensity noise. Furthermore, the noise floor of our technique can be further enhanced through methods discussed in the following section.

The principle of the technique is shown in Fig. 3(a). It is well known that two mutually coherent pulses that are temporally close to each other exhibit spectral interference, whose period *Δv* is determined by pulse separation *τ* as *Δv=1/τ*. Using DFT technique, the spectral interferogram can be mapped to time domain waveforms with a mapping ratio of *2πβL*, where *βL* is the group delay dispersion (GDD) of the system. Therefore, real-time interferometry can be realized by measuring the DFT waveform with fast photodetectors and real-time oscilloscopes [28]. By extracting the pulse-separation evolution at each round trip through Fourier transforming the interferogram, we can estimate the relative timing jitter of a dual-comb system through digital signal processing.

The maximum Fourier frequency of the relative timing jitter PSD is limited to half of the $f_{rep}$ due to the Nyquist condition. This frequency is 20 MHz in our system since the repetition rate of the CANDi laser is 40 MHz. The minimum Fourier frequency ($f_{min}$) is related to $f_{rep}$, $\Delta f_{rep}$ and the maximum resolvable pulse separation ($\tau_{max}$) of the technique through,

$$f_{min} = \frac{\Delta f_{rep}}{f_{rep} \times \tau_{max}} \quad (1)$$

$\tau_{max}$ depends on the dispersion (*GDD*) and detection bandwidth (*BW*) of the system according to,

$$\tau_{max} = 2\pi \times GDD \times BW \quad (2)$$

From Eq. (1), $f_{min}$ can be reduced by increasing $\tau_{max}$, increasing $f_{rep}$ and reducing $\Delta f_{rep}$. This implies that, for a fixed $\tau_{max}$, the required $\Delta f_{rep}$ is smaller in fiber lasers ($f_{rep}$ typically in MHz) compared to microresonators ($f_{rep}$ typically in GHz) for obtaining similar $f_{min}$. From Eq. (2), increasing $\tau_{max}$ requires a large bandwidth photodetector or larger dispersion. In this work, $f_{min}$ of 1 kHz is realized with a $\Delta f_{rep}$ of 0.72 Hz, a GDD of 230 ps$^2$ and a photodetector bandwidth of 12.5 GHz.

The experimental setup for relative timing jitter measurement is shown in Fig. 3(b). A small portion of the combs are monitored using an optical spectrum analyzer (OSA) while the rest of the laser energy are combined using a 50:50 coupler. A polarization controller is installed in one of the arms to align the polarization of two pulses. The combined pulses travel through a 10 km long spool of HI1060 fiber, providing a GDD of 230 ps$^2$, followed by an Ytterbium-doped fiber amplifier (YDFA) stage for amplification. The amplified interferogram signal is detected using a fast photodetector (12.5 GHz bandwidth) and a high-speed real-time oscilloscope (20 GHz bandwidth). It should be noted that the actual energy launched into the DFT fiber is only ~25 pJ (~1 mW average power) to avoid any nonlinearity, which is even lower than many low-energy dual-comb lasers. Hence, this technique is suitable for dual-comb lasers with broad range of pulse energies.

The data processing of the recorded time domain waveform includes a series of steps including selection of a wavelength range containing the high-contrast interferogram, calibration of the interferogram envelope, application of Hanning window and removing nonlinear fringe phase resulting from higher order dispersion during DFT using Hilbert transformation [29], [30]. Finally, a fast Fourier transform (FFT) is applied on this data where the peak position of the FFT result reveals the pulse separation. Repeating these steps for each round trip reveals the evolution of pulse separation. Fig. 3(c) shows a sample interferogram evolution during 80000 laser roundtrips. Fig. 3(d) shows the evolution of pulse separation after processing the interferogram in Fig. 3(c). The deviation from a linear fitting on the pulse separation evolution gives us the relative timing jitter noise. More details can be found in supplementary information (section 1). The effect of nonlinearity on the measurement is further discussed in the supplementary information (section 2).

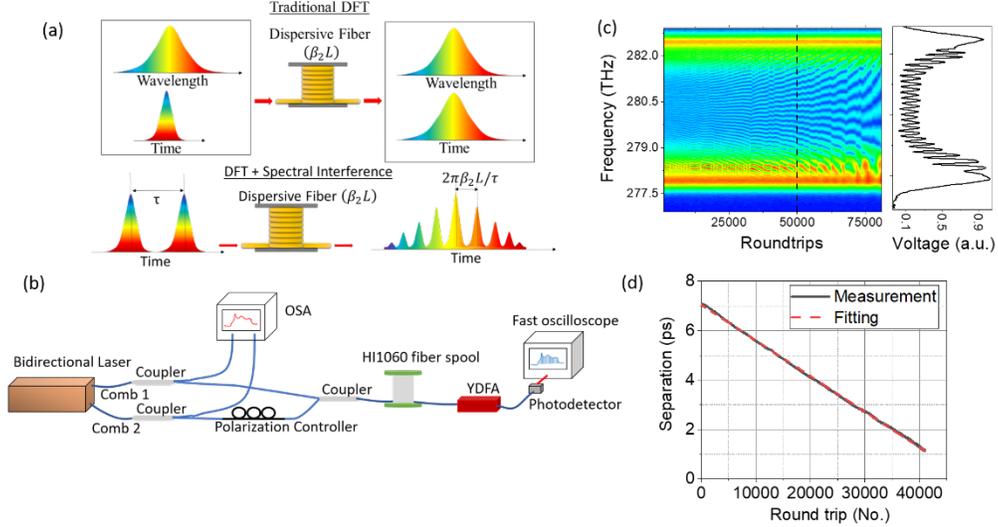

Fig. 3. (a) Principle of traditional DFT and combining DFT and spectral interference. (b) Experimental setup for measuring relative timing jitter in dual-comb lasers using DFT based real-time spectral interferometry. (c) A sample interferogram evolution during 80000 laser roundtrips with each step in the plot corresponding to 1000 roundtrips and a sample interference fringes recorded by the fast oscilloscope. (d) Evolution of pulse separation with roundtrip.

## 5. Relative Timing Jitter Noise of CANDi

Before using the real-time interferometry to measure the relative timing jitter of the CANDi laser, we use a Mach-Zehnder interferometer (MZI) setup to estimate the resolution of our measurement system. One comb from the laser goes through the MZI setup creates a delayed copy of itself and the two pulses then propagate through the rest of the DFT setup. Fig. 4(a) shows the PSD of the system noise (red). The noise floor is about $1.6\times10^{-7}$ $fs^2$/Hz, 5 times lower than the current state of the art by Sandro et al., ($8\times 10^{-7}$ $fs^2$/Hz) [21]. The corresponding noise integrated from 1 kHz to 20 MHz is 1.8 fs, which represents the precision of our DFT system (Fig. 4(b)). The white noise floor is attributed to the 8-bit oscilloscope digitization noise. By replacing the high-loss (~15 dB) dispersive fiber with chirped fiber Bragg grating to enhance the SNR of the detected interferogram and using data acquisition card with larger bit depth, the noise floor can be further reduced, and sub-fs resolution is practically achievable. In addition to the precision, we also measure the accuracy of the technique using time-domain interferometry and the error is only 0.1%. The details of this experiment is explained in the supplementary information (section 3). We also find that the estimated noise floor doesn't change with pulse separation. The dependence of pulse separation in estimating the noise floor using MZI setup is also discussed in supplementary information (section 4).

The black trace in Fig. 4(a) depicts the measured PSD of the relative timing jitter noise of CANDi laser. The integrated relative jitter noise from 1 kHz to 20 MHz (Nyquist frequency) is approximately 39 fs (Fig. 4(b)).

In [14], it has been shown that the timing jitter of the CANDi laser is RIN limited. Here we further show that the relative timing jitter is also pump RIN limited. In order to understand the nature of relative timing jitter in the CANDi laser, we modulate the power of the pump laser of the CANDi laser at different frequencies and the single sideband (SSB) phase noise of individual laser and the relative timing jitter PSD are measured. Fig. 4(c) shows an example measurement with the pump modulation frequency of 4 kHz. The strong CMNR in CANDi laser is evident from the ~12dB difference in noise levels in Fig. 4(c). Apart from the expected peak at the modulation frequency on the phase noise spectrum, we also observe a peak on the relative timing jitter PSD at the modulating frequency. This indicates that the RIN-induced relative timing jitter cannot be suppressed by the CMNR of the CANDi laser and is an uncommon mode noise. Hence, pump RIN could be the major contributor towards the relative timing jitter of the dual-comb CANDi system. To confirm this, the RIN induced relative timing jitter is calculated (supplementary information section 5). The comparison of RIN induced relative timing jitter (green) with the measured relative timing jitter (black) is shown in Fig. 4(d). The reasonable match between the two curves confirms that pump RIN is the dominating factor in the relative timing jitter of CANDi.

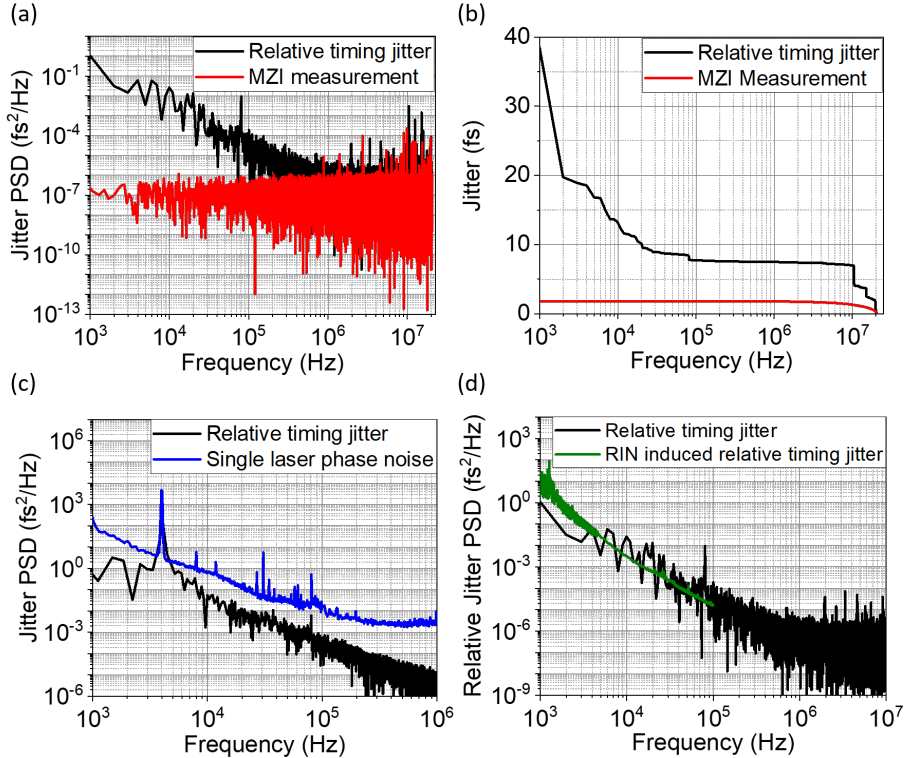

Fig. 4. (a) Measured relative timing jitter results of CANDi laser (black) and MZI experiment (red). (b) Corresponding integrated jitter noise. (c) Measured single direction laser phase noise (blue) and relative timing jitter PSD (black) with pump modulation at 4 kHz. (d) Comparison of the RIN induced relative timing jitter (green) and the measured relative timing jitter (black).

## 6. Coupling Mechanism between RIN and Relative Timing Jitter

In order to minimize the RIN-induced relative timing jitter, it is important to understand the mechanisms contributing to the coupling between pump power change and $\Delta f_{rep}$. For this purpose, we first investigate the coupling mechanisms between pump power change and $f_{rep}$ of both directions. The major coupling mechanisms in a fiber OFC include intensity induced: center frequency shift, change of spectral bandwidth, change of resonant gain and change of self-steepening [31]. Depending on the cavity design, different effects can dominate over the others. The intensity-induced change in the repetition frequency in the case of a true soliton [31] and dispersion-managed solitons [32] have been studied in detail. Here we experimentally identify the major contributors to the intensity-induced repetition rate change in the CANDi laser. Considering all four coupling mechanisms mentioned above, the normalized repetition rate can be written as [31],

$$\frac{1}{f_{rep}} = \beta_1 + \omega_\Delta \beta_2 + \frac{1}{2}\omega_{RMS}^2 \beta_3 + \frac{g}{\Omega_g} + \frac{\mu A^2 \delta}{\omega_0} \qquad (3)$$

where $\beta_n$ are the frequency derivatives of lumped linear fiber propagation constant at gain peak. The terms in the order are group velocity round trip time, spectral shift, TOD, resonant contribution from Yb gain and self-steepening term respectively. $\omega_\Delta = \omega_c - \omega_0$, is the spectral shift of the carrier from gain peak frequency, $\omega_0$ and $\omega_{RMS}$ is the root-mean-square spectral width of the pulse. $g$ and $\Omega_g$ is the gain and gain bandwidth respectively. $\mu$ is the correction term due to modal shape and $\delta$ and $A$ are the nonlinear coefficient multiplied by propagation length and peak electric field of the pulse respectively where $A^2\delta$ is the nonlinear phase change accumulated in the fiber laser. Considering the large pulse width in all-normal dispersion (ANDi) lasers, the self-steepening term can be neglected. Therefore, the rate of change of repetition rate with pump power can be written as,

$$\frac{df_{rep}}{dP} = -f_{rep}^2 \times \left\{\beta_2 \frac{d\omega_\Delta}{dP} + \frac{\beta_3}{2}\frac{d\omega_{RMS}^2}{dP} + \frac{1}{\Omega_g}\frac{dg}{dP}\right\} \qquad (4)$$

To directly measure the magnitude of $df_{rep}/dP$, we apply a triangle wave modulation to the pump power and measure the modulated repetition frequency using frequency counter. Identifying the contributions of different terms on the right side of Eq. (4) to this measured $df_{rep}/dP$ requires estimating several physical quantities. This estimation is detailed in the supplementary information (section 6). The calculated contribution of each of these terms along direction 1 and direction 2 are shown in Fig. 5(a) and Fig. 5(b) respectively. In the CANDi laser, spectral shift term is the largest contributor to the pump power induced timing jitter and spectral bandwidth change is the least contributing term. The difference in spectral shift term along the two directions is due to the asymmetry of the cavity laser structure which results in an asymmetric gain distribution. This gives rise to different gain center frequency shift for a given pump power change which eventually leads to different spectral shift in the two directions. The comparison between the combination of these individual terms and the direct measurement of $df_{rep}/dP$ are shown in Fig. 5(c) and a reasonable match is obtained, which confirms the calculated contribution of each term in Fig. 5(a) and Fig. 5(b). It should be noted that the spectrum evolves considerably inside the ANDi cavity thereby changing the spectral shift and the gain bandwidth terms at different points inside the cavity. The small mismatch in the Fig. 5(c) could be due to the fact that we are measuring the terms only at a single point rather than the average effect. From this, the magnitude of pump power induced relative timing jitter ($d\Delta f_{rep}/dP$) is determined from the difference between the calculated $df_{rep}/dP$ along the two directions. Similar to the pump power induced timing jitter, spectral shift term is the largest contributor to the $d\Delta f_{rep}/dP$ which is followed by the resonant gain term and spectral bandwidth change. This is shown in Fig. 5(d).

Based on these results, we can conclude that the RIN-induced relative timing jitter noise through spectral shift coupled with GDD, i.e., Gordon-Haus effect should be dominant noise

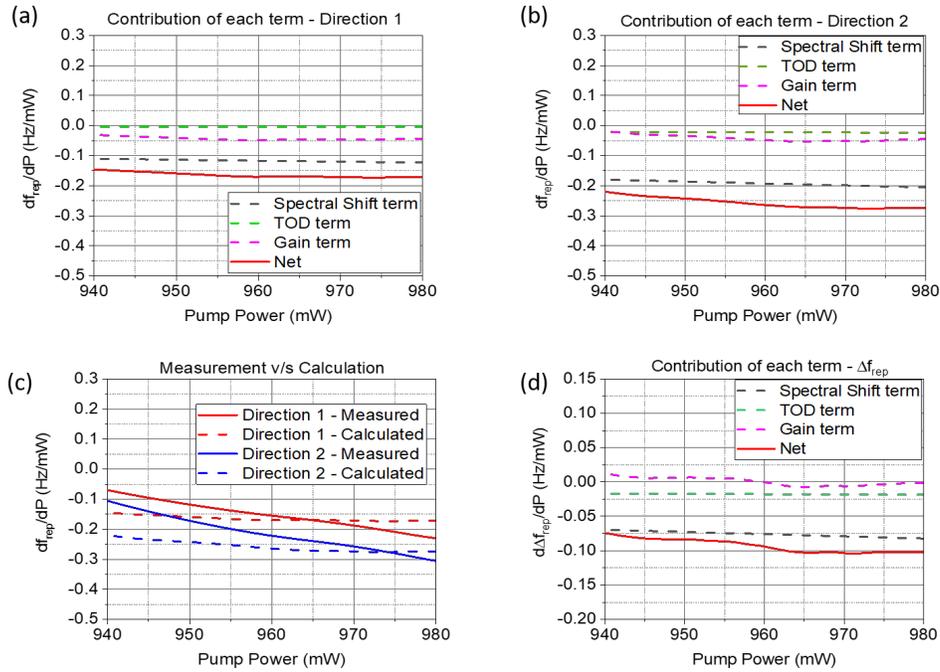

Fig. 5. (a) Individual contribution of each of the three terms in Eq. (4) and the net effect to the dependence of $f_{rep}$ on pump power for direction 1 and (b) direction 2. (c) Comparing the measured dependence of $f_{rep}$ on pump power with calculated value from Eq. (4). (d) Individual contribution of each of the three terms in Eq. (4) and the net effect to the dependence of $\Delta f_{rep}$ on pump power.

source of relative timing jitter in the CANDi laser.

## 7. Discussion

Based on the discussions in the previous sections, it can be deduced that the best way to suppress the relative timing jitter and the individual direction timing error is by reducing RIN-induced timing jitter noise of the laser. From the last section, the most straightforward way to weaken it is reducing the Gordon-Haus effect by lowering the GDD in the laser cavity. However, this will reduce the pulse energy in CANDi laser and negatively impact its application in nonlinear dual-comb systems.

An alternative approach is to reduce the intensity-dependent spectral shift. Therein, such spectral shift can arise from nonlinear effects including self-steepening (SS), Raman self-frequency shift and combined effect of self-phase modulation (SPM) and TOD, as well as from gain filtering effect [32]. Detailed analysis and simulation are performed to evaluate the contribution of each term (supplementary information section 7) and the results are shown in Table 1 along with the experimentally measured center frequency shift. The Raman self-frequency shift is not considered in the simulation due to the opposite sign of its spectral shift compared to the measurement.

Table 1. Simulated magnitude of center frequency shift from different nonlinear effects along with experimentally measured overall shift.

These results shows that the contribution from the SPM + TOD term is negligible and thus

| NONLINEAR EFFECTS CONSIDERED | CENTER FREQUENCY SHIFT (THz/W) |
|---|---|
| SS | 0.007 |
| SPM+TOD | -1.1e-9 |
| SS+SPM+TOD | 0.01 |
| EXPERIMENTALLY MEASURED | 0.18 |

cavity GDD linearization is not an effective solution in minimizing the intensity-dependent spectral shift. Since the net spectral shift from the nonlinear effects (SS + SPM + TOD) is still an order of magnitude lower than the experimentally measured spectral shift, it suggests that the gain filtering effect could be the major source of intensity-dependent spectral shift. A possible way to control the gain filtering effect would be to tune the center frequency by tuning the filter wavelength to operate at a flat region of the gain curve. As discussed in section 6, pump power induced relative timing jitter in CANDi laser is mainly due to the difference in spectral shift along the two directions. This is due to asymmetric gain distribution caused by the asymmetry of the cavity laser structure. Hence, another possible way to suppress the relative timing jitter would be by making the cavity structure and gain distribution more symmetric. Another technique is adopting narrower spectral filters to minimize intensity-dependent spectral shift. However, filter bandwidth needs to be carefully optimized since narrower bandwidth can result in unfavorably higher ASE induced timing jitter due to longer pulse duration [8]. Last but not least, RIN-induced relative timing jitter noise can be reduced by using a pump with better RIN properties or suppressing the RIN of the existing pump using a PLL system.

## 8. Conclusion

In summary, we enhance the state-of-the-art CANDi fiber laser pulse energy from 1 nJ to 8 nJ and introduce a novel reference-free relative timing jitter characterization technique that enables an in-depth analysis on CANDi fiber laser's relative timing jitter characteristics. The measurement noise floor reaches an unprecedentedly low level of $1.6 \times 10^{-7}$ $fs^2$/Hz, and the corresponding measurement precision integrated from 1 kHz to the Nyquist frequency of 20 MHz is only 1.8 fs, currently limited by the 8-bit oscilloscope digitization noise. We show that the cavity length is a parameter that can be used to fine tune the average repetition rate without changing the repetition rate difference or relative timing jitter of CANDi for dual-comb applications. We measure the CANDi's integrated relative timing jitter to be 39 fs [1 kHz, 20 MHz], which is mainly limited by the Gordon-Haus effect induced by the pump RIN. Therefore, improving the pump RIN and minimizing the intensity-dependent spectral shift will further lower CANDi's relative timing jitter. Our study has provided a general guideline for further improving the performance of CANDi fiber laser as well as other single-cavity dual-comb lasers under free-running condition, which will facilitate the development of high-performance compact dual-comb systems and benefit various important fields such as remote sensing, health care and environmental monitoring.

**Funding.** This research was funded by the Office of Naval Research (N00014-19-1-2251) and the National Science Foundation (ECCS 2048202).

**Acknowledgments.** This research was funded by the Office of Naval Research (N00014-19-1-2251) and the National Science Foundation (ECCS 2048202).

**Disclosures.** The authors declare no conflicts of interest.

**Data availability.** Data underlying the results presented in this paper are not publicly available at this time but may be obtained from the authors upon reasonable request

**Supplemental document.** See Supplement 1 for supporting content.

# Unveiling the relative timing jitter in counter propagating all normal dispersion (CANDi) dual-comb fiber laser: supplemental document

**николай Prakash,**[1] **Shu-Wei Huang,**[2] **and Bowen Li**[1,3]

*Department of Electrical, Computer, and Energy Engineering, University of Colorado, Boulder, Colorado 80309, USA*
[1]*These authors contribute equally*
[2] *ShuWei.Huang@colorado.edu*
[3] *Bowen.Li@colorado.edu*

## 1. DFT Data Processing

Measuring relative timing jitter using the DFT based real-time interferometry involves post processing the interferogram signal. A sample portion of the recorded time domain waveform is shown in Fig. S1(a). The time domain waveform is segmented to create a roundtrip evolution map of the interferogram as shown in Fig. 3(c) in the main text. A sample interferogram is shown in Fig. S1(b). From each interferogram, the wavelength range containing the high-contrast interferogram is selected and the interferogram envelope is calibrated using the pulse spectrum envelope. This is followed by application of Hanning window and the resultant interferogram is shown in Fig. S1(c). The nonlinear fringe phase resulted from higher order dispersions during DFT is removed using Hilbert transform. This data processing essentially removes the influence of wavelength dependent GDD and guarantee a linear wavelength-to-time mapping in the DFT process [1], [2]. Finally, a fast Fourier transform (FFT) is applied on the interferogram data, and the peak position of the FFT result reveals the pulse separation. A sample FFT result is shown in Fig. S1(d). Repeating these steps on the interferogram for each consecutive round trip reveals the evolution of pulse separation, which is shown in Fig. 3(d) in the main text. The deviation of the separation evolution from its linear fitting represents the relative timing jitter noise.

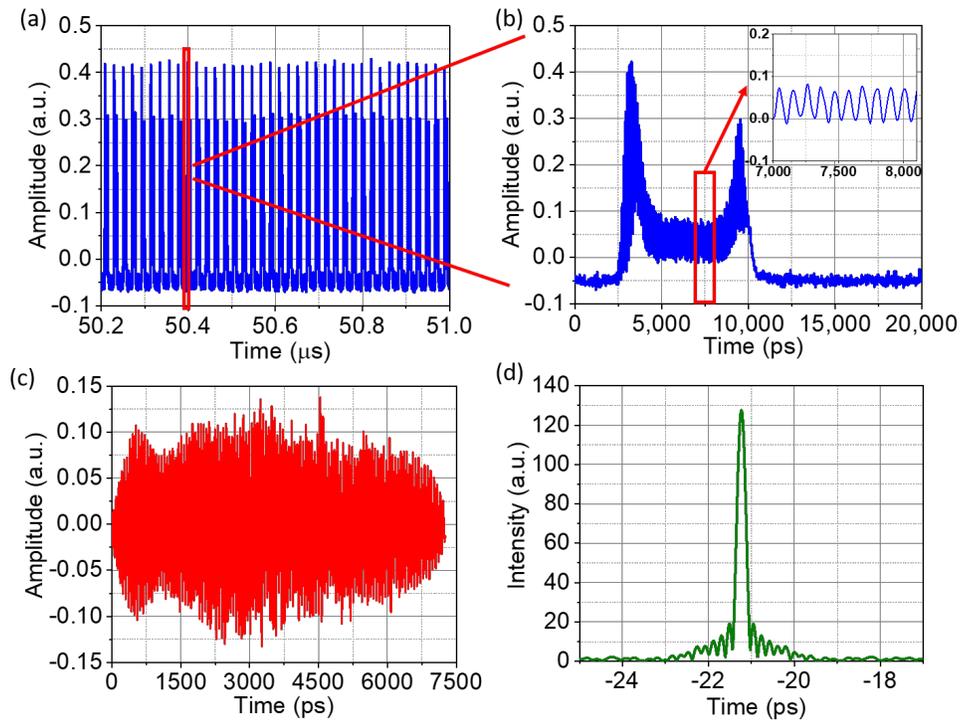

Fig. S1. (a) A section of the sample time domain DFT interferogram waveform recorded using the fast oscilloscope. (b) A sample interferogram in the time domain waveform. (c) A sample interferogram after envelope calibration and windowing operation. (d) A FFT spectrum of the Hilbert transformed interferogram

## 2. Effect of Nonlinearity in the DFT Fiber

The nonlinear interaction between the pulses results in distorted (non-sinusoidal) spectral fringes as shown in the figure below (Fig. S2 (a)). The distortions become more prominent when the pulses are closer to each other. However, this distortion is easily observed on the oscilloscope and results in harmonics peaks in Fourier domain (Fig. S2 (b)). The effect of nonlinearity can be suppressed by reducing the input power into the fiber. A reduced input power measurement result is shown in Fig. S2 (c) and (d). As the input power is reduced, the harmonics peaks are suppressed and the FFT peak gets narrower increasing the accuracy of pulse separation retrieval. In the results presented in the main article, we have made sure that spectral fringes are sinusoidal and are not affected by the nonlinearity by reducing the input average power to <1 mW (< 24 pJ).

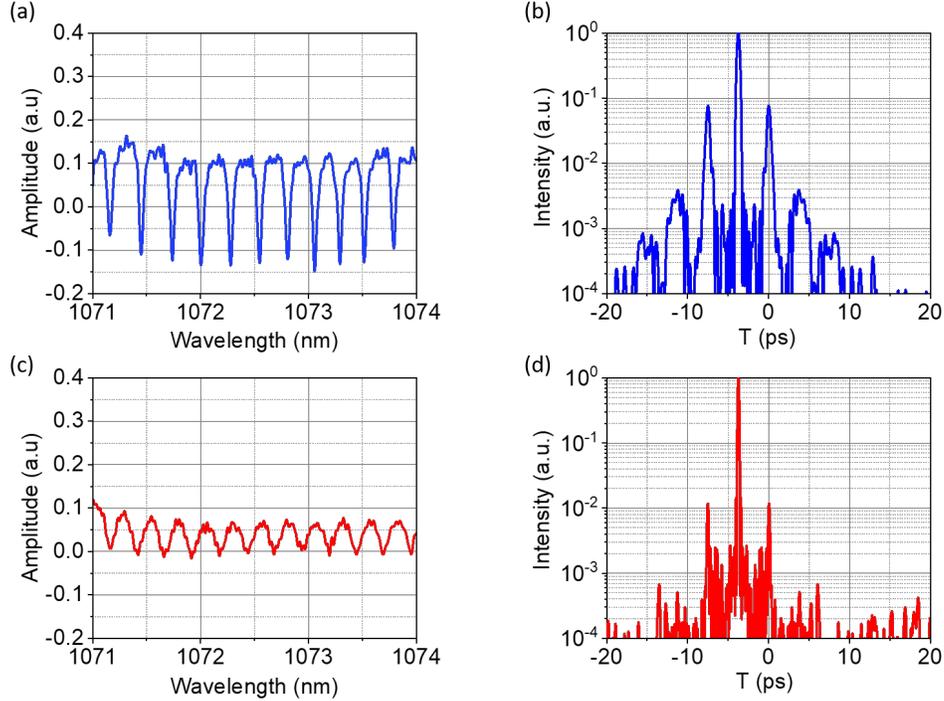

Fig. S2 Effect of nonlinearity on the spectral interferometry. (a) The distorted spectral fringe for a high input power into the DFT fiber. (b) Fourier transform of (a). (c) Spectral fringes when the input power to the DFT fiber is reduced. (d) Corresponding Fourier transform of (c). The pulse separation in these figures is 3.73 ps. Note that the Fourier transform traces are plotted in log scale along vertical axis.

## 3. Accuracy of Pulse Separation Retrieval using Spectral Interferometry and DFT

Apart from the precision, verifying the proposed relative timing jitter measurement method's accuracy is equally important. Since the existing timing jitter measurement techniques have difficulty obtaining the same level of precision, it is difficult to directly validate the method's accuracy using these techniques. Hence, we performed an experimental validation of the presented technique's accuracy using time-domain interferometry. The soul of the proposed technique lies on the accuracy of the pulse-separation evolution retrieval using DFT based spectral interferometry. So, we experimentally compared the accuracy of pulse separation retrieval using the spectral interferometry and DFT technique with the well-established time-domain interferometry using a frequency stabilized continuous wave (CW) laser (NKT Photonics Koheras ADJUSTIK). We used one of the pulses from the dual comb and created a delayed copy of it using a motorized delay stage. We then retrieved the same pulse separation using both DFT based spectral interferometry and time-domain interferometry, respectively. We repeated this operation for several pulse separations and the comparison between the two results are shown in figure below (Fig. S3 (a)). As seen from below, the DFT results match well with the separation obtained from time-domain interferometry. The slope of the linear fit of the DFT method is 0.999 while the expected ideal slope is 1 (i.e., if DFT method exactly matches with the time-domain interferometry, the slope should be 1). The error in the slope of DFT measurement is only 0.1 %. This shows that the technique of pulse separation retrieval using spectral interferometry and DFT and hence, the RTJ estimation are accurate. We have already shown the Mach-Zehnder interferometer (MZI) measurement results confirming the precision of our technique. Hence, the proposed relative timing jitter measurement method is well validated.

## 4. Dependency of Noise Floor on Pulse Separation

As mentioned in the main text, the noise floor of the relative timing jitter measurement using real-time interferometry was measured using a Mach-Zehnder interferometer (MZI) setup. One comb from the laser goes through the MZI setup and creates a delayed copy of itself. The two pulses then propagate through the rest of the DFT setup and the retrieved pulse separation is used to estimate the noise floor of the system. In order to verify that the noise floor estimation is independent of the pulse separation used, we measured the noise floor of the present RTJ measurement technique for three different pulse separation. The results are shown below in Fig. S3 (b). As seen from the results, the estimated noise floor of the technique is almost the same for different pulse separations. Note that the dashed line is the moving averaged curve of each noise PSD denoting the noise floor and all the three dashed lines overlap at $\sim 1.6 \times 10^{-7}$ fs$^2$/Hz.

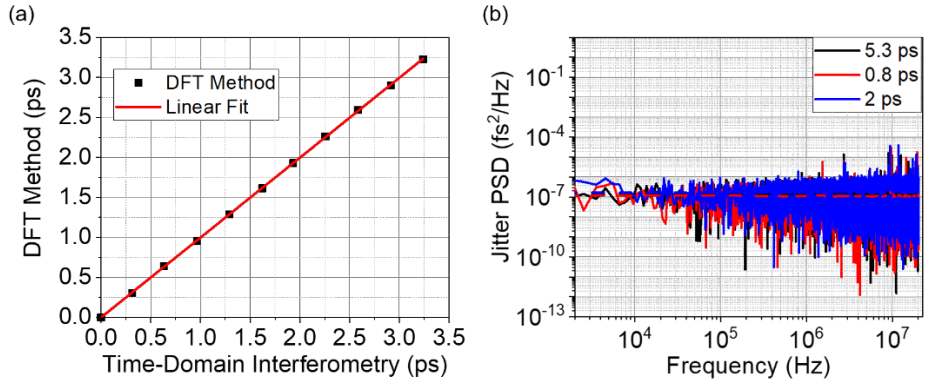

Fig. S3 (a). Comparing the pulse separation measured using DFT and time-domain interferometry. The error in slope of the DFT measurement is only 0.1%. (b) The noise floor measured for three different pulse separations. The dashed line is the moving average of each curve denoting the noise floor.

## 5. RIN Induced Relative Timing Jitter Derivation

The RIN induced timing jitter noise can be directly calculated based on the dependence of $f_{rep}$ on pump power ($df_{rep}/dP$). The power spectral density for the phase noise is related to the timing jitter noise through the relation [3],

$$\Delta S_\varphi(f) = (2\pi f_{rep})^2 \times \Delta S_{\Delta t}(f) \quad (S1)$$

where $S_\varphi$ is the phase noise power spectral density and $S_{\Delta t}$ is the timing jitter noise spectrum.

The contribution of fluctuation of round trip time to the timing jitter noise is given according to [3],

$$\Delta S_{\Delta t}(f) = \left(\frac{1}{2\pi f T_{rt}}\right)^2 \times S_{T_{rt}}(f) \quad (S2)$$

where $f$ is the frequency, $T_{rt}$ is the round trip time and $S_{T_{rt}}$ is the spectral power density of fluctuation of round trip time. Since fluctuation in round trip time can be expressed in terms of pump power ($P$) as

$$\Delta T_{rt} = \frac{dT_{rt}}{dP} \times \Delta P \tag{S3}$$

$S_{Trt}$ can be expressed in terms of pump RIN ($S_I$) as,

$$S_{Trt}(f) = \left(\left(\frac{1}{f_{rep}}\right)^2 \times \frac{df_{rep}}{dP}\right)^2 \times S_I(f) \tag{S4}$$

From Eq. (S1), (S2), (S3) and (S4),

$$\Delta S_\varphi(f) = \left(\frac{\frac{df_{rep}}{dP}}{f}\right)^2 \times S_I(f) \tag{S5}$$

The above equation gives the RIN induced phase noise. Since the coefficient $df_{rep}/dP$ is different along both the directions, the RIN induced relative timing jitter (RTJ) can be calculated from the RIN induced phase noise along the two directions as,

$$\Delta S_{RTJ}(f) = \Delta S_\varphi(f)_{Direction2} - \Delta S_\varphi(f)_{Direction1} \tag{S6}$$

The pump RIN and laser RIN are measured using a FFT analyzer and are shown in Fig. S4. Note that the laser RIN is similar along both directions and hence, laser RIN along direction 1 is shown in Fig. S4. As seen below, the major noise peaks in pump RIN up to 1 kHz are also visible in laser RIN, after which there is a gain response filtering effect. The similar magnitudes (~20 dB) of the peak at 100 kHz in both pump and laser RIN confirms that the laser RIN is pump RIN limited but with a gain response filtering effect. Therefore, laser RIN and dependence of $f_{rep}$ on laser power ($df_{rep}/dP_L$) can be used to predict pump RIN induced relative timing jitter with the effect of gain response filtering. The coefficient $df_{rep}/dP_L$ for the laser directions 1 and 2 are 951 Hz/W and 1051 Hz/W respectively.

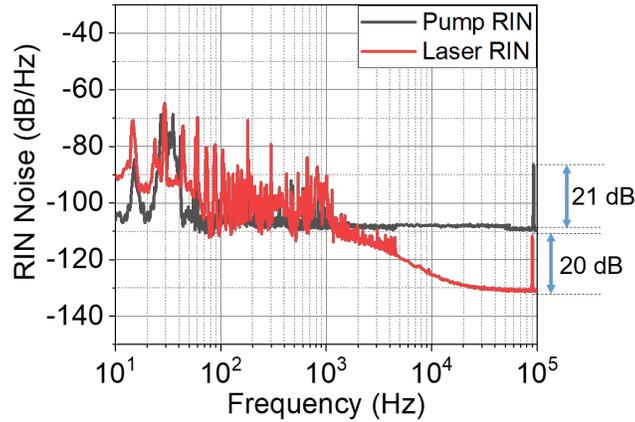

Fig. S4. PSD of the laser RIN and pump RIN.

## 6. Estimating the RIN to Timing Jitter Noise Coupling Coefficient

Identifying the contributions of the different coupling mechanisms (right side of Eq. (4) in the main text) causing the dependence of $f_{rep}$ on the pump power requires estimating several physical quantities. We estimate the dispersion terms by measuring the group velocity dispersion (GVD) and third order dispersion (TOD) of the fibers used in the system through white light interferometry. The measured $\beta_2$ and $\beta_3$ are about 0.103 ps$^2$ and 3.948e-5 ps$^3$ respectively. The change in spectral shift and RMS spectral width with pump power ($d\omega_\Delta/dP$ and $d\omega^2_{RMS}/dP$) are calculated by recording the output spectra of the CANDi at different pump power using an OSA. The center frequency and RMS spectral width are estimated from the recorded spectra and are fitted with a 2$^{nd}$ order polynomial. The fitted curves are shown in Fig. S5. $d\omega_\Delta/dP$ is estimated from the center frequency change by assuming a constant gain peak frequency for Yb doped fiber.

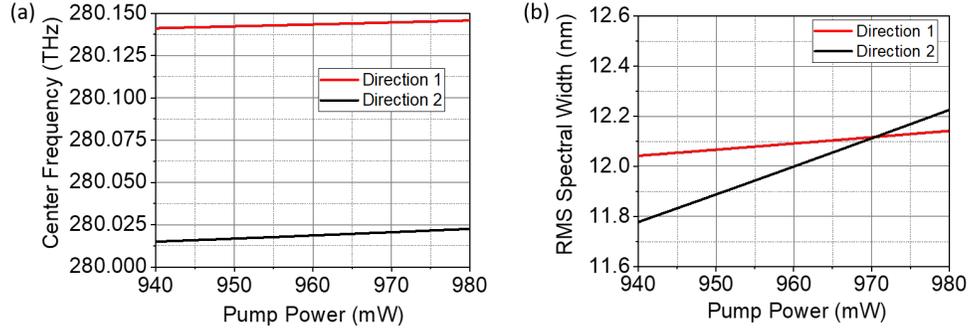

Fig. S5. Dependence of (a) center frequency and (b) RMS spectral width on pump power for the two directions of CANDi estimated from the output spectra.

The gain term is estimated from the measured change in loss in the CANDi laser with pump modulation. The main losses considered are the polarization dependent loss at the polarizing beam splitter and the frequency dependent loss at the band pass filter. The polarization dependent loss is calculated by measuring the change in output power and intracavity power with pump modulation while the loss induced by the band pass filter is estimated based on the measured output spectra of CANDi laser at different pump power and the Gaussian spectral profile of the filter centered at 1070 nm with a bandwidth of 2.62 THz. These measurements are shown in Fig. S6. At steady state, using the fact that gain is equal to loss, we can estimate $dg/dP$ from the change in loss. Note that the gain bandwidth is assumed to be same as the filter bandwidth which is also the shortest bandwidth in the system.

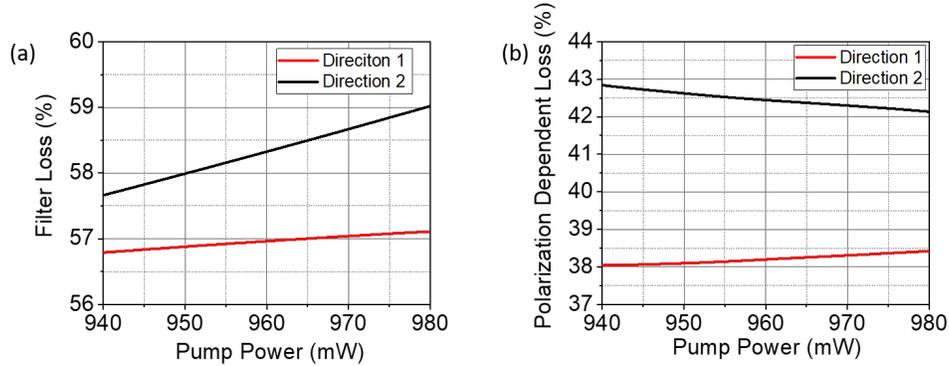

Fig. S6. Dependence of (a) filter loss and (b) polarization dependent loss on pump power for the two directions of CANDi.

Experimentally, $df_{rep}/dP$ is estimated by applying a 0.1 Hz triangle wave modulation to the pump power and recording the modulated repetition frequency using frequency counter. This is shown in Fig. S7 (a).

## 7. Simulating Effects of Nonlinearity on the Intensity-Dependent Spectral Shift

The nonlinear effects considered for simulation are self-steepening (SS), Raman self-frequency shift and combined effect of self-phase modulation (SPM) and TOD. The SS results in blue shift while Raman will induce a red shift. The sign of spectral shift due to SPM +TOD depend on the sign of TOD. From experiments, the net spectral shift is found to be blue shifted (Fig. S5). Hence, Raman self-frequency shift cannot be the major contributor to the nonlinear spectral shift. In order to compare the effects of SS and SPM+TOD term, simulations are performed for the Gaussian pulse propagation through the CANDi laser setup using parameters similar to experiments and considering the different nonlinear effects. Fig. S7 (b) shows the center frequency as a function of pump power for SS, SPM + TOD and SS + SPM + TOD. Note that the sign of TOD is positive. The center frequency shift considering only the SS term is 0.007 THz/W while that due to SPM + TOD is only -1.1e-9 THz/W. The net spectral shift due to SS + SPM + TOD is 0.01 THz/W.

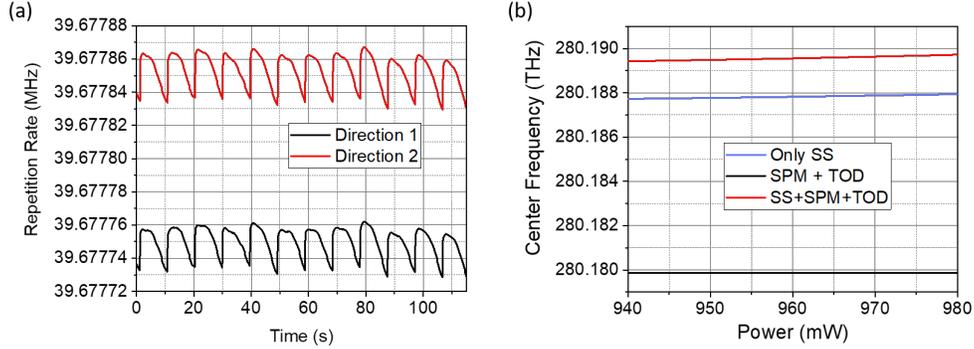

Fig. S7. (a) Modulation of $f_{rep}$ using a triangle wave modulation of pump power. (b) Simulation results showing the center frequency as a function of pump power considering SS, SPM + TOD and SS + SPM + TOD.

## 8. Effect of Pulse Interaction Position on Relative Timing Jitter Measurements

In order to confirm that the relative timing jitter measurement results do not depend on the interaction position of the two counter-propagation pulses in the laser cavity, we have measured the relative timing jitter of CANDi laser for different interaction positions in the cavity. We achieve the different interaction positions by changing the fiber length difference (by increment of 1 m) on the two arms of fiber coupler, where the two outputs of the CANDi laser are combined. This is illustrated in Fig. S8 (a). As seen from the results below (Fig. S8 (b)), the relative timing jitter PSDs are relatively overlapping for the different interaction positions in the cavity. This means that the pulse interaction position inside the cavity does not affect the relative timing jitter behavior. Hence, it is not necessary to determine the exact position of pulse interaction. This also highlights the advantage of minimized cross-talk between the two pulses in a bidirectional mode-locking laser scheme.

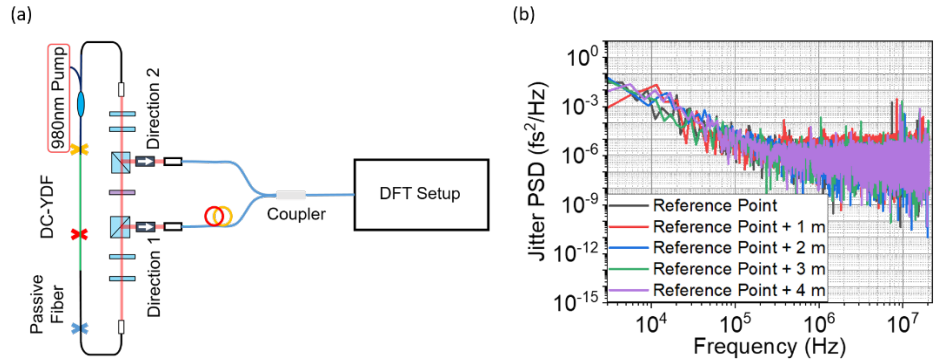

Fig. S8. Effect of interaction position of the pulses in the cavity on the relative timing jitter PSD measured using the presented technique. (a) Experimental setup and (b) Relative timing jitter PSD results. The interaction position inside the cavity for each additional fiber length added is illustrated by 'X' and the different colors imply different fiber path length.

## Supplemental Document References